\documentclass[conference]{IEEEtran}

\ifCLASSINFOpdf
  \usepackage[pdftex]{graphicx}
\else
  \usepackage[dvips]{graphicx}
\fi
\usepackage{tabulary,url,paralist}
\usepackage{tikz}
\definecolor{specialblue}{RGB}{0,124,232}

\usepackage{multirow}

\usepackage{fancyhdr}

\usepackage{geometry}
\fancypagestyle{plain}{
  \fancyhead[L]{2015 Models and Technologies for Intelligent Transportation Systems (MT-ITS)
\\3-5. June 2015. Budapest, Hungary}     
  \fancyfoot[L]{ISBN 978-963-313-142-8 @ 2015 BME }

}

\begin{document}
\pagestyle{plain}

\title{A State-of-the-art Integrated Transportation Simulation Platform\vspace{-2ex}}

\author{\IEEEauthorblockN{Tiago Azevedo, Rosaldo J. F. Rossetti, Jorge G. Barbosa}
\IEEEauthorblockA{Artificial Intelligence and Computer Science Lab\\
Department of Informatics Engineering\\
Faculty of Engineering, University of Porto, Portugal\\
\{tiago.manuel, rossetti, jbarbosa\}@fe.up.pt}
}

\maketitle

\thispagestyle{plain}

\begin{abstract}
Nowadays, universities and companies have a huge need for simulation and modelling methodologies. In the particular case of traffic and transportation, making physical modifications to the real traffic networks could be highly expensive, dependent on political decisions and could be highly disruptive to the environment. 
However, while studying a specific domain or problem, analysing a problem through simulation may not be trivial and may need several simulation tools, hence raising interoperability issues.
To overcome these problems, we propose an agent-directed transportation simulation platform, through the cloud, by means of services. We intend to use the IEEE standard HLA (High Level Architecture) for simulators interoperability and agents for controlling and coordination. Our motivations are to allow multiresolution analysis of complex domains, to allow experts to collaborate on the analysis of a common problem and to allow co-simulation and synergy of different application domains.
This paper will start by presenting some preliminary background concepts to help better understand the scope of this work. After that, the results of a literature review is shown. Finally, the general architecture of a transportation simulation platform is proposed.
\end{abstract}


%

\section{Introduction}

Nowadays, universities and companies all around the world have a huge need for simulation and modelling methodologies. The objectives are varied, but simulation is widely used for decision making and what-if analysis, as well as for performance optimisation, testing, training, and so forth. In the particular case of traffic and transportation, making physical modifications to the real traffic networks could be highly expensive, dependent on political decisions and could be highly disruptive to the environment. Therefore, simulation is broadly used in such scenarios.

However, while studying a specific domain or problem, analysing a problem through simulation may not be trivial and very often requires several simulation tools, with different resolutions and domain perspectives, hence raising interoperability issues. Thus instead of helping, simulation could be a headache! Transportation problems usually are complex and fall within this category of problems. 

To date, there are not many solutions for traffic that make full use of the intelligent agent concept. However, the multi-agent system metaphor has become recognised as a convenient approach for modelling and simulating complex systems~\cite{moya2007towards}. Also, it has grown enormously not only for being applied to traffic but also to transportation in general terms~\cite{bazzan2013review}.

With the recent evolutions in Cloud Computing and Software-as-a-Service (SaaS), there is a new paradigm where simulation software is used in the form of services. Indeed, such evolutions have been more significantly seen in the business world with Information Technology solutions moving to the SaaS paradigm~\cite{sharif2010s}. So, Simulation Software-as-a-Service (SimSaaS) is very beneficial to better exploit the huge amount of platforms and storage that simulation needs \textit{per se} - and Cloud Computing is able to provide such resources. This way, researchers do not need to have the simulation software installed on their own computers or access servers which host this kind of software. Besides, they have new computing environments and methodologies for software development over the Internet. 

To address issues arising in this novel perspective the main goal of this paper is to propose an agent-directed transportation simulation platform, through the cloud, by means of services. It is intended to use the IEEE standard HLA (High Level Architecture) for simulators interoperability and agents for controlling and coordination. To do so, it is necessary to build the body of knowledge needed to develop such a platform. This paper's objective is to present the current state of the art in the field and, with that, also to present the general architecture of the intended platform.

The motivations of the platform are to allow multiresolution analysis of complex domains, to allow experts to collaborate on the analysis of a common problem and to allow co-simulation and synergy of different application domains. It is expected to fulfil three main contributions. Firstly, a technological contribution because one will have a cloud-based simulation platform for transportation using HLA and agents where simulations are offered in the form of services. Secondly, a scientific contribution since it will enable the collaboration among experts of the Modelling \& Simulation (M\&S) field with the agent-directed paradigm. Finally, an applied contribution with an agent-oriented platform for scientific simulation, through the cloud, by means of services. Basically, it will be a virtual laboratory.

This paper will start to present some preliminary background concepts regarding M\&S, the agent-oriented paradigm, HLA and cloud for a better understanding of the scope of this work. After that, the results of a literature review concerning SimSaaS in the Cloud, HLA and services, and agent-directed SimSaaS is shown. Finally, the general architecture of a transportation simulation platform is proposed.

\section{Preliminary Background}

\subsection{M\&S in Traffic and Transportation}
In the context of M\&S, a system is defined as a collection of entities, for example people or machines that act and interact together toward the accomplishment of some logical end~\cite{schmidt1970simulation}.

There no unique definition of M\&S in the literature, depending on each domain and scientific field. According to the Merriam-Webster Online Dictionary (http://www.merriam-webster.com/), simulation is ``the imitative representation of the functioning of one system or process by means of the functioning of another'' or, in other words, the ``examination of a problem often not subject to direct experimentation by means of a simulating device''. So, in simulation we have not only the idea of representation but also the idea of an experimentation without the direct intervention of a human.

Modelling is basically an abstract and simplified representation of a system. It is similar to the system, but simpler~\cite{robinson2008conceptual}, as it should be an approximation to the real system with the most relevant features, but simple enough to be understandable. As~\cite{sargent2005verification} points, a good model is a judicious trade-off between realism and simplicity.

As it can be seen, simulation of a system is mainly the execution of a model. Indeed, simulation is widely used for decision making and what-if analysis, as well as for performance optimisations, testing, training, and so forth.

Most of the times it is not possible to make direct experimentation with the actual systems, and it is necessary to make simplifications using modelling. In this case, there is the question of whether it actually reflects the system, but analysing this issue is outside the scope of this paper. If the model is simple enough, it is possible to study the system using an analytical solution through, for example, calculus, algebra and probability theory. Otherwise, if the system is highly complex, simulation is performed using computational means. 

In the particular case of traffic and transportation, making physical modifications to the real traffic networks could be highly expensive, dependent on political decisions and could be extremely disruptive to the environment. Therefore, M\&S is broadly used in such scenarios and related tools can provide better and more concise data for analysis.

\begin{figure}[h!]
\centering
\includegraphics[width=.9\linewidth]{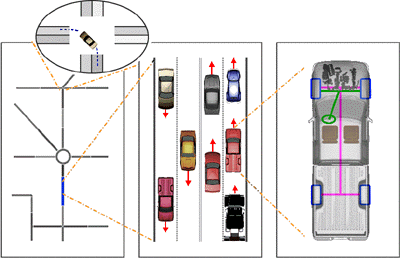}
\caption{The major four categories of modelling in traffic (from left to right): macroscopic, microscopic, nanoscopic (mesoscopic within the circle)~\cite{krajzewicz2002sumo}}
\label{fig:traffic_models}
\vspace{-0.9em}
\end{figure}

There are several ways to model traffic depending on the level of detail in which they describe the traffic dynamics. Figure~\ref{fig:traffic_models} illustrates the major four categories of modelling in traffic, namely macroscopic, mesoscopic, microscopic and nanoscopic.

Macroscopic models describe traffic in terms of flows (the number of vehicles that pass through a certain road per hour) or densities, without considering their entities such as vehicles. Therefore, these models are good to analyse large or complex networks.

Unlikely macroscopic models, microscopic models detail both the behaviour of each entity and their interactions, with each other as well as with the network. For that, these models incorporate vehicle behaviour rules such as acceleration, breaking, lane changing, and so forth.

Mesoscopic models fill the gap between macro and micro models. They normally describe traffic entities at a high level of detail, whereas their behaviour and interaction are briefly described. In these models, vehicles can be grouped in packets, which are routed throughout the network and treated as one single entity.

A new trend in traffic simulation though is the nanoscopic model which extends the capabilities of three basic components of microscopic simulation: vehicle modelling, vehicle movement modelling, and driver behaviour modelling~\cite{dia2008nanoscopic}. For example, in the case of traffic, it intends to describe the components of a vehicle. Attempts at integrating these different perspective in a more transparent way through modelling has already been reported in the literature~\cite{rossetti1999} \cite{ferreira2008}.

\subsection{Agent-directed Simulation}

The introduction of \textit{intelligent demons} (called intelligent agents today) to control simulation experiments was introduced by the MISS Hungarian Center~\cite{Javor2012}. This notion of \textit{intelligent agent} may not be consensual, mainly because of the difficulty of defining what \textit{intelligence} is. However, we consider an agent as an autonomous and proactive computational entity whose rational process is based on concepts such as Knowledge, Belief, Intention, Commitment, Goal, Desire and Emotion.

The intelligent agent concept brings a genuine metaphor to represent autonomous entities as it is equipped with sensors and effectors as well as with reasoning and decision-making abilities. Thereby, we have entities with high-level communication skills who also provide endless possibilities for system coordination and controlling. Otherwise, system coordination and controlling would be reduced to automated scripts, which do not bring so many advantages.

To date, there are not many solutions for traffic that make full use of the intelligent agent concept. However, the multi-agent system approach has become recognised as a convenient approach for modelling and simulating complex systems~\cite{moya2007towards} and has grown enormously not only applied to traffic but also to transportation in general terms~\cite{bazzan2013review}. Nevertheless, just a few simulation tools truly support the concept of agents and multi-agent systems in traffic simulation; MATSim-T~\cite{balmer2009matsim} and ITSUMO~\cite{bazzan2010itsumo} are good examples to be mentioned. Also, besides agent-based traffic simulation, the MAS-Ter Lab platform introduces the concept of expert agents in charge of the simulation analysis process~\cite{rossetti2007}.

The agent-oriented paradigm has a panoply of associated concepts, being important to define them and point out what is our perspective on those definitions.

The agent metaphor encompasses some other concepts, that~\cite{yilmaz2007agent} unify in the Agent-Directed Simulation paradigm. The authors indicate that the paradigm consists of three distinct, yet related areas that can be grouped under two categories: Simulation for Agents (agent simulation) and Agents for Simulation. The first is about simulation of systems that can be modelled by agents, that is, the simulation model is an agent or, in other words, we are simulating agent systems. The latter can still be grouped under two categories, namely agent-based simulation and agent-supported simulation, as follows.

\begin{compactitem}
  \item \textbf{Agent-based simulation} is the use of agent technology to generate model behaviour or to monitor generation of model behaviour. The perception feature of agents makes them pertinent for monitoring tasks. Agent-based simulation is useful for having complex experiments and deliberative knowledge processing such as planning, deciding, and reasoning.
  \item \textbf{Agent-supported simulation} deals with the use of agents as a support facility to enable computer assistance by enhancing cognitive capabilities in problem specification and solving. Hence, agent-supported simulation involves the use of intelligent agents to improve simulation.
\end{compactitem}

A lot of researchers do not take into account the contribution of agents to simulation. Thus, in such cases \textit{agent simulation} and \textit{agent-based simulation} are seen as the same principle. In this work, it is adopted the same perspective in which the two principles are seen indistinguishably. 

\subsection{HLA}

Parallel and distributed simulation (PADS) relies on partitioning the simulation model across multiple execution units. Each execution unit manages only a part of the model and handles its local event list, but locally generated events may need to be delivered to remote execution units~\cite{DAngelo2014320}.

Distributed simulation facilitates the reuse of heterogeneous simulation systems but has issues regarding interoperability of simulators. HLA is an IEEE software standard developed to provide a common technical architecture for distributed M\&S, trying to provide the structural basis for interoperability among simulators.

In HLA, every participator of the simulation is called federate, and these federates can interact with each other within a federation. The baseline components of HLA include (1)~Federate Interface Specification, (2)~Framework and Rules, and (3)~Object Model Template (OMT) Specification.

The Federate Interface Specification has the services which federates can use for communication~\cite{ieee2010hlaFederate}. This communication between simulators is managed by a Run-Time Infrastructure (RTI). In order to make possible the interaction between federates and the RTI, there is the concept of ambassador. Basically, federates communicate with the RTI using its ambassador as an interface.

The Framework and Rules of HLA are the set of rules which must be obeyed in order to ensure the proper interaction within a federation. There are five rules for federates and also five rules for federations, detailed in~\cite{ieee2010hlaRUules}.

The OMT describes the format and syntax of the data exchanged among federates. Thus, it defines the object template data that all simulation unit needs to use in order to exchange data with each other~\cite{ieee2010hlaOMT}.

\begin{figure}[!ht]
\centering
\includegraphics[width=.9\linewidth]{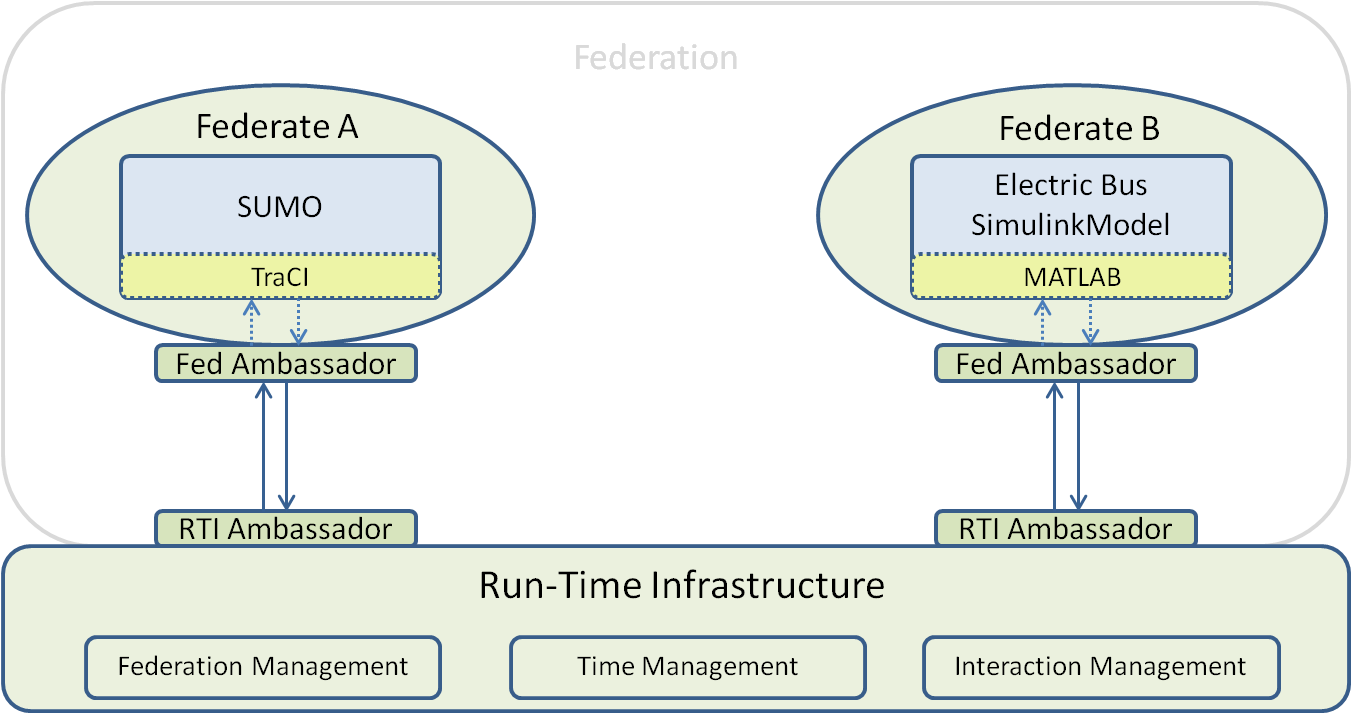}
\caption{ HLA's Functional Architecture with two different simulators \cite{macedo2013integrated}}
\label{fig:hla_macedo}
\vspace{-0.9em}
\end{figure}

Figure \ref{fig:hla_macedo} depicts all referred concepts instantiated for the work described in~\cite{macedo2013intelligent}. In this work it was presented a distributed architecture for electric bus powertrain simulation within a realistic urban mobility context. There, the communication between \textit{Federate A} ambassador and SUMO simulator~\cite{behrisch2011sumo} is performed through SUMO's API, TraCI. In this sense, whenever simulation data is required, the RTI ambassador performs calls to federate ambassador that communicate with SUMO through TraCI.
In a similar way, the communication between \textit{Federate B} ambassador and Simulink~\cite{perrotta2012potential} model is performed through MATLAB, the Simulink's API.

\subsection{The Cloud Computing Paradigm}
In 1969, Leonard Kleinrock~\cite{kleinrock2005vision} said: ``As of now, computer networks are still in their infancy, but as they grow up and become sophisticated, we will probably see the spread of `computer utilities' which, like present electric and telephone utilities, will service individual homes and offices across the country''. After so many years, this is becoming true, and computing is even more managed and delivered in the form of traditional utilities such as water, electricity, gas and telephony. Cloud Computing is helping to leverage this utility vision.  

Consequently, Cloud Computing has become another buzzword and more and more work is being done in the field. Nevertheless, there are dozens of different definitions for Cloud Computing and there is little consensus on that~\cite{geelan2009twenty}.

Despite this vision, it is possible to differentiate cloud computing from grid computing. In this research, we adopt the definition of Cloud Computing provided by The National Institute of Standards and Technology~\cite{mell2011nist}, as it is an Institute responsible for standards and we did not see until now any other standard definition:

\textit{Cloud computing is a model for enabling ubiquitous, convenient, on-demand network access to a shared pool of configurable computing resources (e.g., networks, servers, storage, applications, and services) that can be rapidly provisioned and released with minimal management effort or service provider interaction.}

As also shown in~\cite{armbrust2010view}, there are three aspects that are new in cloud computing: (1)~the appearance of infinite computing resources available on demand, (2)~the elimination of an up-front commitment by cloud users, and (3)~the ability to pay for the use of computing resources on a short-term basis as needed.

Cloud Computing is a fresh and on-going hot topic not only for the industry but also among academics. Indeed, even recently new buzzwords emerged from Cloud Computing, trying to extend it even further: Fog Computing~\cite{bonomi2012fog} and Cloud 2.0~\cite{miluzzo2014m} are examples. For a wider state-of-the-art perspective on Cloud Computing with important research directions there are several recent sources to analyse \cite{gonzalez2015cloud} \cite{zhang2010cloud} \cite{Sun2014134} \cite{Abdelmaboud2015159} \cite{he2014state} \cite{abolfazli2015mobile}.

\section{Literature Review of Cloud SimSaaS}

\subsection{SimSaaS in the Cloud}

Simulation Software-as-a-service (SimSaaS) is a relatively new paradigm where simulation software is used in the form of services. Thanks to the significant attention on the Cloud computing paradigm and taking advantage of the several resources provided, it is possible to set simulations into the cloud. This way, it is possible to offer simulations through the cloud by means of services. 

In our research, it turned clearly evident that SimSaaS is a very recent topic. Just five papers are prior to 2011 and the first three do not specifically use the term SimSaaS, vaguely mentioning \textit{simulation} and \textit{web}. Besides, there is an increase number of work over the years.

Although the amount of papers found about SimSaaS is not quite big, it is wide in which concerns application domains. For example, in the biomedical domain there is a system devoted to simulations of electromagnetic field inside the human body~\cite{Sawicki20121190}. In crowd and pedestrian M\&S there is also one work proposing a method based on a distributed architecture with simulation in the cloud, with the authors indicating that there is a lack of automation and integration of tools for crowd M\&S~\cite{Wang:2015:SSM:2723553.2723554}. As~\cite{Sliman2013611} suggest, research dissemination methods suffer from a major drawback as they do not allow publishing simulation code and scripts along with the published papers. So, the authors demonstrate an ongoing project in which scientists can openly share their underlying code and data.

A lot of other specific application domains use SimSaaS, mostly in the cloud, and more specifically work regarding ontology learning~\cite{Wang2014179}, traffic and transportation~\cite{Harri2010},scheduling parallel discrete event simulation jobs~\cite{Liu2012} and a cloud simulation in manufacturing~\cite{Taylor201489}, just to mention a few.

In addition to these specific domain works on SimSaaS, there are also generic ones. For example,~\cite{Tsai:2011:SSS:2048370.2048381} proposed a SimSaaS framework incorporating multi-tenancy architecture and scalability for simulation, also presenting a simulation run-time infrastructure.

\begin{figure}[h!]
\centering
\includegraphics[width=.9\linewidth]{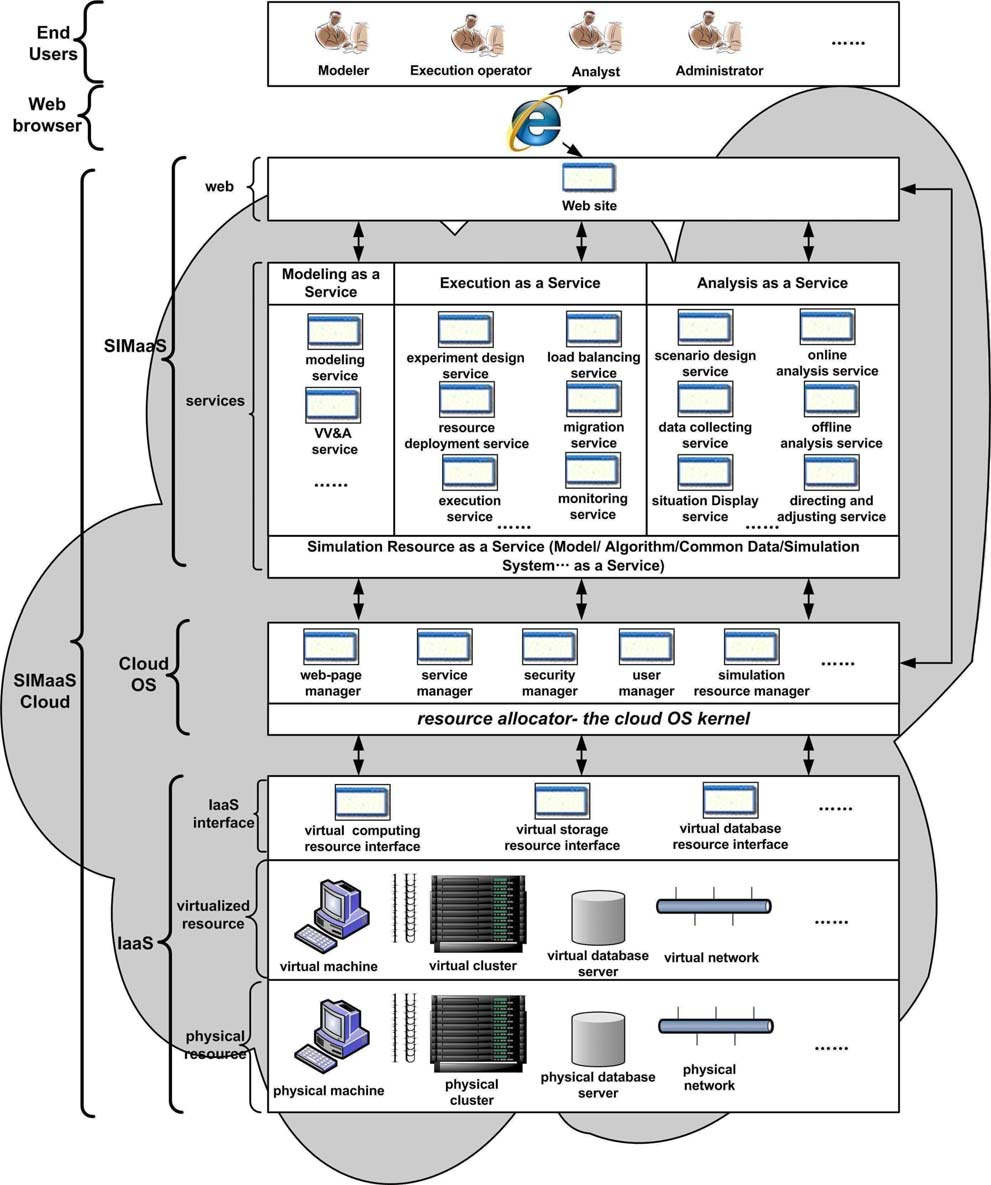}
\caption{ The architecture of the Cloud Simulation~\protect\cite{Liu201271}}
\label{fig:csim_arch}
\vspace{-0.9em}
\end{figure}

As~\cite{Liu201271} indicate, pioneers such as Richard M.Fujimoto, Bohu Li, Gabriele D'Angelo and so forth, made considerations on the Cloud Simulation but none has given an overall picture of Cloud Simulation to its full extent. Consequently, the authors propose the general architecture of a Cloud Simulation, which is a SaaS type cloud. This architecture is illustrated in Figure~\ref{fig:csim_arch}.

In general terms, it is possible to verify the infrastructure in the bottom of the architecture, where virtualisation plays a vital role. Before reaching the end users, which can be very diverse, the architecture shows a great detail in the specification of the offered services. Authors divide services into three self-explanatory groups: Modeling as a Service, Execution as a Service and Analysis as a Service. During the provision of these services, available simulation resource can be reused with the aid of the Simulation Resource as a Service. Managing and connecting the baseline infrastructure and services there is the so-called Cloud Operating System.

\subsection{HLA and services}
While specifically discussing SimSaaS, there is almost no clearly references about interoperability among simulators. However, HLA, the current standard for simulator interoperability, has been used in many different works, such as in agent-based simulations. In the context of this paper, it is important to show how it is possible to extend HLA to enable simulations offered in the form of services.

A first approach can be seen in~\cite{Xie:2005:SPH:1069810.1070184}. In this paper, the authors propose a framework to extend the HLA to support Grid-wide distributed simulation. Here, a remote proxy acts on behalf of the federate in interacting with the RTI. It hides the heterogeneity of the simulators, simulators' execution platforms, and how the simulators communicate with the RTI.

In a work done by~\cite{Bocciarelli:2013:SAF:2675983.2676156}, it is introduced \textit{HLAcloud}, a model-driven and cloud-based framework to support both the implementation of a Distributed simulation system from a SysML (Systems Modelling Language dialect of UML) specification of the system under study and its execution over a public cloud infrastructure. 

Maybe, the most relevant paper found is the one that presents a possible way for HLA to be integrated with a Service Oriented Architecture (SOA) in the context of a smart building project~\cite{Dragoicea:2012:IHS:2310096.2310200}. The paper discusses the design of an HLA federate for the inclusion of a service oriented smart building controller in the simulation loop. It is important to refer that the Simulation Manager module is a service wrapper on top of the RTI that exposes access to the RTI's federation management via a RESTful API. It deals with the creation, initialisation, deletion, starting, stopping, and execution of simulations.

Still in that paper, the authors cite~\cite{Wang:2008:SHL:2367656.2367672} who presented a comparison between HLA and SOA concluding that:
\begin{compactitem}
\item HLA has good interoperability, synchronisation and effective and uniform information exchange mechanism between the communicating components (federates), but lacks several features of web services, such as the integration of heterogeneous resources, web-wide accessibility across firewall boundaries;
\item SOA benefits from loose coupling, component reuse and scalability but lacks a uniform data exchange format and time synchronisation mechanisms;
\item The combination of HLA and SOA can extend the capabilities of the two technologies and thus enable integrated simulated and real services.
\end{compactitem}

\subsection{Agent-directed Simulation-as-a-service}
Agent-directed simulation and, more specifically, agent-supported simulation is used in a huge variety of fields. Nevertheless, in the field of Simulation-as-a-service, just a few examples exist, and even fewer concern the cloud.

Albeit there is such a lack of work regarding agents and cloud, work such as~\cite{Yilmaz01092006} already mentioned the importance of agents to simulation (even in gaming) by exploring the relationship of software agents with simulation and games.

An agent-supported simulation is seen in~\cite{Guo:2012:TGE:2346616.2346654}, where \textit{handling agents} were used for composition of simulation services with different time and event granularity. In this case, the composition consisted of wildfire and weather simulation services, but no cloud was used.

Elsewhere~\cite{Tolk:2010:UFA:2433508.2433550}~and~\cite{Tolk:2011:MTI:2431518.2431551} suggest that most current simulation interoperability standards are insufficient as they focus exclusively on information exchange to support the federation of solutions without providing the necessary introspection. HLA is a bit more flexible, as the information to be exchanged itself is not standardised (it only says how to structure the data), but the focus remains on information that can be exchanged with a system. So, the formal approach to simulation interoperability (using agent-supported simulation) tries to solve this problem.

The only work found that truly implements agents in the cloud indicates that cloud computing can speed up significantly agent-based M\&S to facilitate more accurate and faster results, timely experimentation, and optimisation~\cite{Taylor:2014:TCC:2693848.2693884}. However, the many different clouds, cloud middleware and service approaches make the development of agent-based M\&S in the cloud highly complex.

Already in 2004~\cite{Yilmaz01092006} pointed that there were limitations of existing federated simulation environments in supporting dynamic model and simulation updating. HLA federation development, for instance, requires complete specification of object models and information exchanges before the simulation run begins. They also observed and argued that the lack of machine processable formal annotations describing the behaviour, assumptions and obligations of federates is a fundamental roadblock. After a decade, it seems that this problems still remain almost the same.

\subsection{Summary}

SimSaaS is a trendy term with high potential to grow more. However, there are problems with the term: there is a lack of automation and integration of tools in M\&S~\cite{Wang:2015:SSM:2723553.2723554}, and research dissemination methods suffer as they do not allow publishing simulation code and scripts along with the published papers~\cite{Sliman2013611}. But not everything is bad! Although the amount of papers found about SimSaaS is not quite big, it is wide in what concerns application domains.

Simulator interoperability is a very explored subject in general, but when it comes to SimSaaS, almost nothing focus on this. Indeed,~\cite{Yilmaz:2014:PFR:2693848.2694204}, in their panel about the future of research in M\&S, authors refer to the distribution of SimSaaS in the cloud as a future research topic.

HLA is another term referenced a lot in the literature since the first complete version (HLA 1.3) was published in 1998. However, again, there is a few works regarding extension of HLA to allow simulation services in general and in the cloud. Nevertheless, HLA solely has some disadvantages~\cite{Yilmaz01092006}\cite{Tolk:2010:UFA:2433508.2433550}.

Although cloud computing can speed up significantly agent-based M\&S to facilitate more accurate and faster results~\cite{Taylor:2014:TCC:2693848.2693884}, agents are not an exception and there is an absence of work putting agents in the cloud to support SimSaaS. Nonetheless, an example was shown where it was possible to develop an agent-supported simulation to bear out SimSaaS~\cite{Shao:2009:SSE:1639809.1639859}.

Summing up, it is possible to see lots of gaps in the literature concerning SimSaaS, SimSaaS in traffic and transportation, SimSaaS in the cloud, HLA in the cloud, solutions to HLA restrictions, agents to support SimSaaS and agents in the cloud. So, these will be the front lines of this work! Knowing that there are yet a few works regarding SimSaaS and synergies of all these gaps, this paper aims to bridge them by proposing an architecture of a transportation simulation platform in the cloud.

\section{A seamlessly integrated transportation simulation platform}

\begin{figure}[h!]
\centering
\includegraphics[height=.25\textheight]{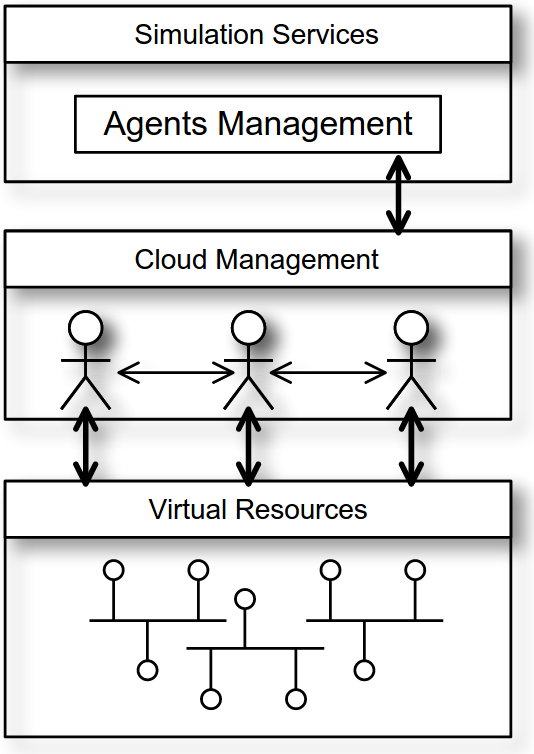}
\caption{Generic architecture of the proposed platform}
\label{fig:general_arch}
\vspace{-0.9em}
\end{figure}

According to the gaps previously found in the literature review, we propose an integrated transportation simulation platform regarding every term (SimSaaS, HLA and Agent-directed simulation), as it will offer simulation in the form of services, using HLA for interoperability of simulators and agents for collaboration. The general architecture of such platform will be similar to the one described in Figure \ref{fig:csim_arch}. The main differences to such architecture is that it will be adapted in order to support HLA in the \textit{Virtual Resources} tier and Agents in the \textit{Cloud Management} tier. Figure~\ref{fig:general_arch} shows the generic architecture of the proposed platform considering the three main tiers that differ from the generic cloud simulation.

The scientific community needs collaboration in its pursuit of multidisciplinary achievements. This way, scientific community started to make a first sharing approach by creating public repositories of datasets, which are sustainable, shared and ever-evolving. However, there are other roles/stakeholders that would also benefit from such datasets, such as public decision-makers and the industry alike. Nevertheless, while they are all interested in testing their own algorithms/calculations, just the scientific community generally adopts the philosophy of data sharing. Besides data, we also have processes, methods and plans, which are even less commonly shared.

In this multi-stakeholder and multi-resource philosophy, HLA already supports the inclusion of resources because a federate is sufficiently general to consider not only simulators but also databases, data loggers, and other resources. Therefore, the so-acclaimed agent-directed approach could be used to enable collaboration between experts, sharing not only data but also processes. But, why is this interest in sharing? What is the interest in creating this shared community above the cloud? We believe the answer is simple and consists in generating knowledge and innovation.

Simple sharing methodologies are, for example, personal pages in the platform for each researcher with the created models, similar suggested models, integration of models of other researchers and performed simulations with obtained results. Nevertheless, the agent-directed paradigm could bring better options. For example, some researchers have their own legacy tools, which run some complex and less optimised simulations, producing outputs. It would be much better if these tools were implemented in the platform, and it also had the possibility to deploy agents that act on behalf of the researcher, that initialise these tools and even that generate graphics from the resulting output. It would be like an avatar. In practice, these are services that exist in agents: possibility to start and stop everything, make data collection, make data analysis and even pick the output to serve as the input to other tools.

How the researcher can implement and deploy their own agents? The platform itself in the cloud allows that! As well as there is a methodology like HLA to interoperate simulators, there would exist also a design methodology for each researcher to implement his or her own agent/avatar. Each avatar is characterised by the perceptions, actions and operators that could feature. The cognition is decided by the researcher and is expected to implement his or her own expertise. In summary, this is more than just a scientific environment for empirical science. There is the need to be more than that in order to value stakeholders like public decision makers and industry in general, as they could implement their own simulations in the platform.

There are many fields where such a platform can enhance simulations. For instance, cloud computing is becoming increasingly deep-seated in our lives, and Mark Weiser once said ``The most profound technologies are those that disappear. They weave themselves into the fabric of everyday life until they are indistinguishable from it''~\cite{weiser1991computer}. Indeed, living labs can be supported by simulation, as the people in these cases would be receiving stimuli from other virtual realities, supported by the platform.

This platform may seem too futuristic but nowadays technological advances have all the proper means that allow its implementation. Firstly, the ones concerning the management and control of the cloud itself. The amount of existing tools in the field is huge, but they are mostly commercial. While looking for alternatives, decision comes mostly between OpenStack (http://www.openstack.org/) and OpenNebula (http://opennebula.org/).

The implementation of HLA's RTI in the cloud is also possible. Again, there are several tools but most of them do not implement the full HLA specification, or have limitations for free use. Here, the choice is easier and falls in the only two tools that seem to be still active, namely the PoRTIco project (http://www.porticoproject.org/) and Pitch pRTI (http://www.pitch.se/).

\section{Conclusion}

This paper started to explain its context, motivation and objectives, following clarification of important concepts for a better understanding of the approach herein proposed. Results of the literature review were broadly presented, focusing on the current (and a few) synergies among concepts such as Cloud, HLA and agents in the cloud. It was thus possible to verify the current research level regarding SimSaaS, Cloud Computing, HLA and Agent-directed simulation. 

The lack of published research on coupling these terms suggests an important area needing further investigation, which is explored in our approach. As a result, we shared our vision and general architecture of an agent-directed transportation simulation platform, through the cloud, by means of services. Next steps in our research include the implementation of the specified general architecture, and further investigation of ways to use agents in the cloud, over the RTI tier. It will then be possible to leverage the technological, scientific and applied contributions as pursued in this research. As a first approach to implement such a platform, the example previously mentioned~\cite{macedo2013integrated} should be executed in the cloud. If someone would be capable to execute such a concrete application for the platform without installing anything and yet being able to parametrise it, the platform will immediately prove its benefits.

\section*{Acknowledgement}
This work has been partially supported by MIEIC, Faculty of Engineering, University of Porto.

\bibliographystyle{IEEEtran-modified}
\bibliography{literature}

\end{document}